\documentclass[final,5p,authoryear,twocolumn]{elsarticle}	

\usepackage[british]{babel}
\usepackage{graphicx}
\usepackage{longtable}
\usepackage{amssymb}
\usepackage{color}
\usepackage{rotating}
\usepackage{overpic}
\usepackage{hhline}

\newif\ifreview
\reviewtrue
\reviewfalse
\ifreview
\input{reviewpreamble}
\fi

\usepackage[colorlinks%
	,bookmarks,bookmarksnumbered%
	,breaklinks,hyperindex%
	,linkcolor={blue}%
	,citecolor={blue}%
	,anchorcolor={blue}%
	,pagecolor={blue}%
	,menucolor={blue}%
	,urlcolor={blue}%
	,pdfauthor={Carsten Lemmen <carsten.lemmen@hzg.de>}%
	,pdfsubject={}%
	,pdfkeywords={early agriculture, Neolithic package, pre-Columbian population, adaptation, sociotechnological model},
	,pdftitle={Different mechanisms shaped the transition to farming in Europe and the North American Woodland}
	]{hyperref}%

\newcommand{\executeiffilenewer}[3]{%
\ifnum\pdfstrcmp{\pdffilemoddate{#1}}%
{\pdffilemoddate{#2}}>0%
{\immediate\write18{#3}}\fi%
}

\newcommand{\reftab}[1]{Table~\ref{#1}}

\newcommand{\reffig}[1]{Figure~\ref{#1}}

\newcommand{\comment}[1]{}

\newcommand{\pskm}{\ensuremath{\mathrm{km}^{-2}}}

\newcommand{\science}[2]{\ensuremath{#1\cdot 10^{#2}}}

\journal{Archaeology, Ethnology and Anthropology of Eurasia}

\begin{document}
\begin{frontmatter}
\title{Different mechanisms shaped the transition to farming in Europe and the North American Woodland}

\author[hzg]{Carsten Lemmen\corref{lab:cor}}
\cortext[lab:cor]{Tel +49\,4152\,87-2013, Fax~-2020, Email~carsten.lemmen@hzg.de} 
\address[hzg]{Helmholtz-Zentrum Geesthacht, Institute of Coastal Research, Max-Planck Stra\ss e~1, 21501~Geesthacht, Germany}

\begin{abstract} 
The introduction and emergence of agriculture into Eastern North America and Europe proceeded very differently: it varied in timing, speed, and mechanism---despite similar woodland environments.  To resolve the different subsistence paths, I employ the Global Land Use and technology Evolution Simulator, a numerical model for simulating demography, innovation, domestication, migration and trade within the geoenvironmental context.  I demonstrate how Europe receives a large package of foreign domesticates and converts rapidly. In contrast, Eastern North American trajectories exhibit a gradual transition: hunting-gathering and agropastoralism coexist for a long time, and agriculture is integrated slowly into the existing subsistence scheme.
\end{abstract}


\end{frontmatter}

\section{Emergence of farming in woodlands}

In the Levant and along the Chinese Yellow and Yangtze rivers, the first green revolution occurred between 10,000,\,BC and 8000\,BC when the predominant hunting-gathering life style was replaced by agriculture and pastoralism as the main subsistence style \citep[][]{Willcox2005,Kuijt2002,Londo2006}.  
Subsequently, this new way of actively procuring food resources took hold in almost every society around the globe---at different times, with different speed, and different mechanisms.  Where agriculture had not arisen independently, or where the local domesticates where insufficient to serve as staple crops, agriculture was imported from the founder regions either by prehistoric trade network or by migrants \citep[e.g.,][]{Wen2004,Lemmen2011}.   Prime exemplary regions for non-independent agriculture are Europe and Eastern North America \citep[ENA,][]{Price2001,Smith1997}.  

In most of Europe, the transition to farming occurred between 6000\,BC and 4000\,BC, with a continuous temporal gradient from the southeast (earliest agriculture in Greece and the Balkan, \citealt{Perles2001}) to the northwest (southern Scandinavia, \citealt{Price2003}, and Britainm \citealt{Whittle2007}).  The most prominent example for a large-scale homogeneous agricultural complex is the Linearbandkeramik culture of central Europe \citep{Luening2005bha}.   None of the crops used had been domesticated locally in Europe, but had been imported from the west Asian founder regions; everywhere, the emergence of agriculture was associated with the appearance of ceramic artifacts.
The Neolithic package which was delivered to Europe is comprised of food crops and animals that have their origins in the wider region of the Fertile Crescent \citep{Flannery1973}, it contains wheat, barley, rye, lentils, peas, cattle, sheep, goat, and pigs  \citep{Willcox2005,Luikart2001,Edwards2007,Larson2007,Zeder2008}.

The transition to farming occurred much later in North America, between 1000\,BC and 1000\,AD, a period denoted for ENA as the Woodland stage.  Local domesticates like amaranth and sunflower had been cultivated along a dominating hunting gathering life style, but gradually corn and beans were being introduced from Mexico and became the staple crops.  The Woodland period is characterized by continuous pottery artifacts and the gradual increase of the dependence on agriculture \citep{Anderson2002}.

The success of growing crops is restricted to suitable climatic conditions and sufficient soil and water resources.  With the exception of the dry plateaus along the Western cordillera,  North America and Europe exhibit similar climatic conditions, vegetation types and topography, both with a variety of landscapes and features like large rivers, mountain ranges, plains and rolling hills. The natural vegetation type on the majority of North American and European area is temperate forest \citep{Williams2000,Thompson2000,Cheddadi1997}.

In this study, I employ the Global Land Use and technological Evolution Simulator  (GLUES) which resolves local innovation, migration and cultural diffusion of traits \citep{Wirtz2003gdm,Lemmen2011}.    With GLUES I demonstrate that the model is capable to explain the global onset of agriculture for many world regions very reasonably; here, I restrict my analysis of the global model to North America and Western Eurasia.  I elaborate on the differential processes leading to agriculture in Europe and the Eastern North American Woodland region, and I estimate the pre-Columbian population in the Woodland region.

\section{Material and Methods}
The Global Land Use and Technological Evolution Simulator (GLUES, \citealt{Wirtz2003gdm,Lemmen2009,Lemmen2010,Lemmen2011}) describes the evolution of regional sociocultural traits constrained by biogeographical setting.  Available resources exploited by regional societies are based on regional net primary production, derived for the past by applying  \nobreak{Climber-2}-simulated temperature and precipitation anomalies  \citep{Claussen1999} on the IIASA climatological data base \citep{Leemans1991}, and a bioclimatic limit vegetation model \citep{Lieth1975}. The conceptual model is outlined below, for details on the algorithms used and the mathematical implementation we refer the reader to \citet{Wirtz2003gdm} and \citet{Lemmen2011}.

\subsection{Sociocultural model}

The sociocultural realm is described by three characteristic traits of a population and its demographic density.  The temporal evolution of each trait follows the gradient of increased benefit for success (i.e., growth) of its associated population \citep{Dieckmann1996dtc,Wirtz1996eve}.  The evolution of the distribution of each trait in terms of its moments, not a single realization of the trait values is simulated. Below, I give a short description of the characteristic traits:

(1)~Technology is a trait which describes the efficiency of food procurement and the effectiveness of basic health care.  In particular, it describes the availability of tools and weapons, storage and work organization.  Included in this trait is writing as a means for administration and cultural storage, and transport.
(2)~The share of farming and herding activities (energy or time or manpower) with respect to the total food sector. 
(3)~The number of agropastoral economies available to a regional population; though I do not attribute named plants and animals to each economy, as an example, a number of four is obtained when pig farming, barley and wheat harvesting, and goat herding are present in one population.
With this setup I follow \citet{Shennan2001} by directly coupling population dynamics with the process of cultural evolution.

Model parameters are constrained by requiring that the emergence of agropastoral centers in five founding regions is successfully simulated with appropriate timing as suggested by \citet[][p12]{Smith1997}: these are the Fertile Crescent (7500\,BC), central Mexico (7000\,BC), south China (8500\,BC), north China (7800\,BC), south central Andes, (7000\,BC) and the eastern United States (4500\,BC)\footnote{In this paper, I use present day names to refer to approximate geographic location}.  For each simulation with random parameter variation, a score is calculated considering the spatiotemporal distance to these centers.  The best scoring parameter set from one million simulations was selected for the hindcast study in this paper.

The mathematical model conceptually works as follows:  Within each biogeographically defined region (average size \science{127}{3}\,km$^2$), a Mesolithic hunter-gatherer population is incubated at simulation start.  The Mesolithic population grows with a speed constrained by sociocultural traits, and is limited by natural resources.  As technologies become more powerful, resources are more efficiently exploited and first domesticates are brought under control.  Demographic density and technologically advanced exploitation put pressure on the environment and reduce the usable resources.  Only when the combination of technologies and domesticates outcompetes the hunter-gatherer subsistence with respect to marginal growth rate benefits, a local transition to farming occurs.  Typically, the local transition is completed in less than 200~years, consistent with empirical data \citep[e.g.,][]{Zvelebil1996}.

Trade and migration are vectors of the spread of farming from the founding centers to more distant regions.   The mechanism of trade and migration relies on the concept of influence difference \citep{Renfrew1979edp}, which describes the tendency and direction of exchange.  Migration occurs with speed proportional to length of the contact zone divided by the distance of two neighbors.  Trade exchange is proportional to the difference in the products traded, and is assumed to reach one hundred times farther than migration. \citet{Lemmen2011} recently investigated the importance of trade and migration as vectors of cultural or demic diffusion followed by local adoption in the GLUES model; they found that both processes are consistent with the radiocarbon record in Europe.

For the mathematical model, \citet{Lemmen2011}  introduced the age scale ``simulated time BC'' (sim\,BC) to distinguish between empirically determined age models and the model time scale; I adopt this notation here.  Ideally, sim\,BC should be numerically equal to year BC as calendar year.  GLUES version~1.1.17 was used for this study with the standard 685~world region Linearbandkeramik (LBK) setup \citep{Lemmen2009} and no imposed regional climate fluctuation.  GLUES is free and open source software and can be obtained from \href{http://glues.sourceforge.net/}{http://glues.sourceforge.net/}.

\subsection{Biogeograpic and bioclimatic forcing}

\begin{figure}
\ifreview\else\parbox[c]{0.63\hsize}{\fi%
\includegraphics[width=\hsize]{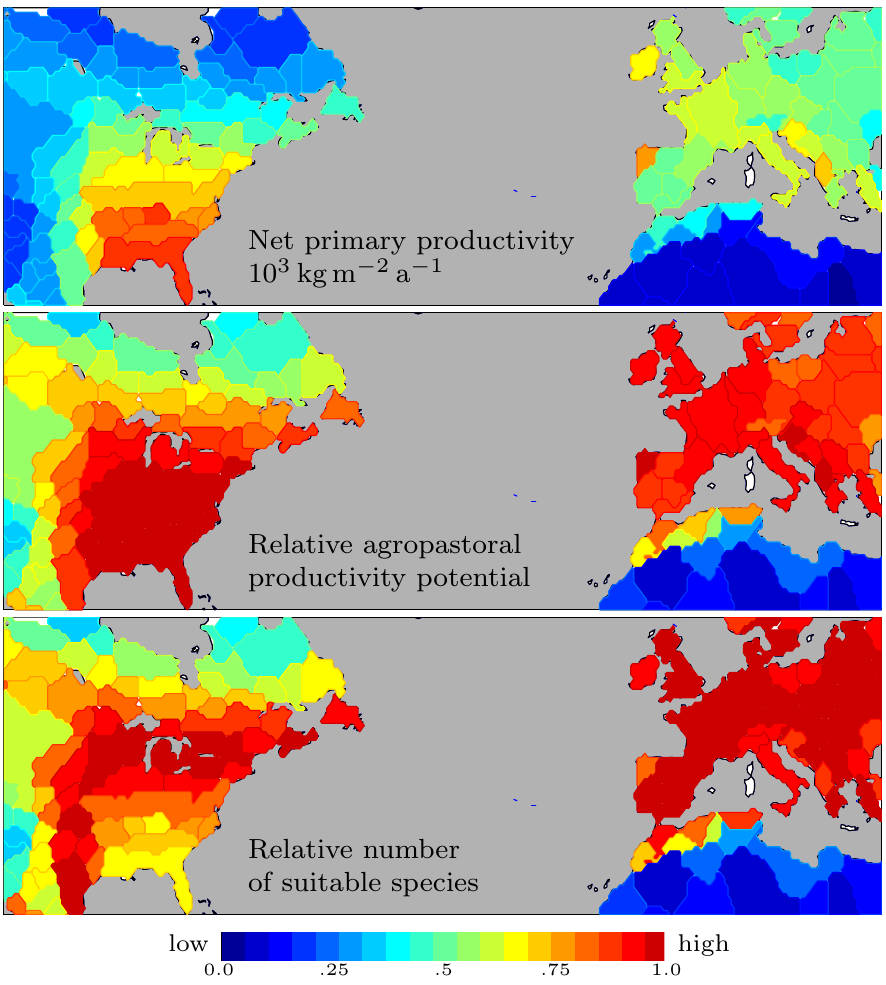}%
\ifreview\else}\hfil\parbox[c]{0.36\hsize}{\fi%
\caption{Biogeographic context for the simulation of sociotechnological evolution with GLUES.  The biogeographic variables productivity (top panel), agropastoral potential (middle panel) and number of suitable species (bottom panel) are derived from reconstructed temperature and precipitation at 1000\,sim\,BC}
\label{fig:map_resources}
\ifreview\else}\fi
\end{figure}

The sociotechnological model is embedded in the geographic and past vegetation context.  In GLUES, vegetation is measured by its net primary productivity (NPP); net primary productivity is derived from a bioclimatic limit model \citep{Lieth1975} on annual mean temperature and annual precipitation.  To obtain the vegetation of the past, I use results from a transient Holocene \nobreak{Climber-2} \citep{Claussen1999} intermediate complexity climate model simulation as anomalies on the IIASA climatological data base for mean monthly values of temperature and precipitation \citep{Leemans1991}.

\reffig{fig:map_resources} shows  available natural resources relevant for the simulated transition to agriculture:  the highest productivity is found in the southeast of North America, intermediate values throughout the eastern and central  United States and throughout Europe, low values in boreal Canada, the Rocky Mountains and the Sahara (top panel in Figure~\ref{fig:map_resources}).  The figure  shows vegetation productivity at 1000\,sim\,BC, which is representative for the entire simulation because changes in the climate data set are so small that the simulated productivity does not vary much during the Holocene. 

Albeit higher in the southeast of North America, the vegetation productivity is in a suitable range (500--1500 grams per square meter and year) in both the ENA and European woodland regions, when it is converted to the relative agropastoral productivity potential  (middle panel in Figure~\ref{fig:map_resources}). With the transfer function defined by \citet[][their Figure~1]{Wirtz2003gdm} for the conversion from net primary productivity to agropastoral productivity potential, this potential is high (above 75\%) for all of Western Europe and the eastern and central United States.

The bottom panel in Figure~\ref{fig:map_resources} shows the relative number of suitable species---which is related to the number of different subsistence economies---in Europe and Eastern North America.  The transfer function from net primary productivity to local species diversity has a narrow optimal range of 400---600 grams per square meter and year \citep[][their Figure~1]{Wirtz2003gdm}, which indicates the open woodland optimum for agriculturally suitable (e.g., annual grasses, grazing herbivores) animal or plant resources. The simulated species diversity is very high throughout Europe; it is very high in the north eastern and the central United States, but only moderately high in the south eastern United States.   To account for the expected higher diversity in large areas versus small areas, already shown almost two centuries ago by \citet{Watson1835}, I employ a concave species--area relationship to estimate from the local resources the continentally available diversity of suitable species \citep{Connor2001,Wirtz2003gdm}.  On a continental scale, the potential suitable species diversity in Eurasia is four times the potential which is to be expected for North America. 

\section{Results}

\begin{figure}
\centering
\includegraphics[width=\ifreview\else0.8\fi\hsize]{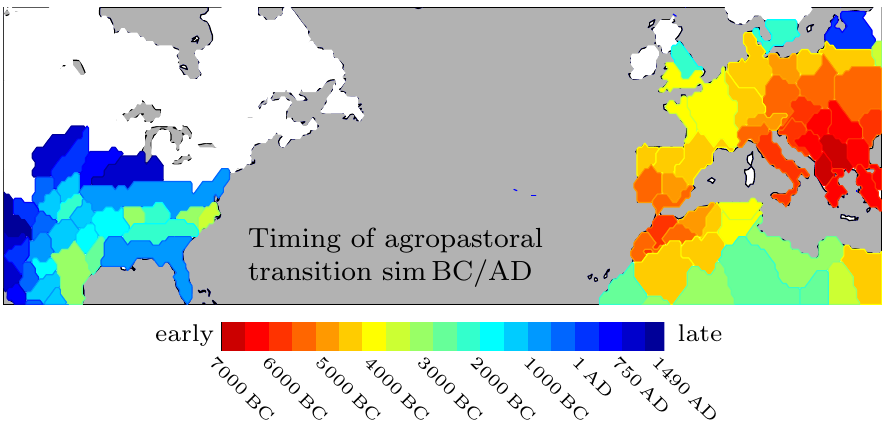}
\caption{Regional timing of the transition to agriculture or pastoralism as the dominant subsistence style, hindcasted with the GLUES socio-technological simulator.  All colored regions have converted to agropastoral subsistence, while white coloring indicates that the model does not predict the transition there by 1490\,AD.}
\label{fig:map_timing}
\end{figure}

Farming and pastoralism as major subsistence styles emerge in the model simulation almost everywhere on the globe by 1490\,sim\,AD.   Hunting and gathering remain important in the boreal zone of Europe and North America  (Figure~\ref{fig:map_timing}), and on large islands like Iceland (not shown) and Ireland.  The timing of the transition to agropastoralism---detected in the model when more than half of the regional population rely on farming or herding---is depicted by colors in Figure~\ref{fig:map_timing}.  There is a  timing offset between Europe and ENA, with most of continental Europe converting to agropastoralism during the period 7000--4000\,{sim\,BC}, and most of ENA converting between 2500\,{sim\,BC} and 1490\,{sim\,AD}.
This  lag in the simulated timing of the onset of agriculture between Europe and Eastern North America is a consequence of the different biogeographic setting of the two subcontinents as discussed by \citet{Wirtz2003gdm}. 

\subsection{European agricultural expansion}

\begin{figure}
\centering
\ifreview\else\parbox[c]{0.53\hsize}{\fi%
\includegraphics[width=\ifreview0.8\else1\fi\hsize]{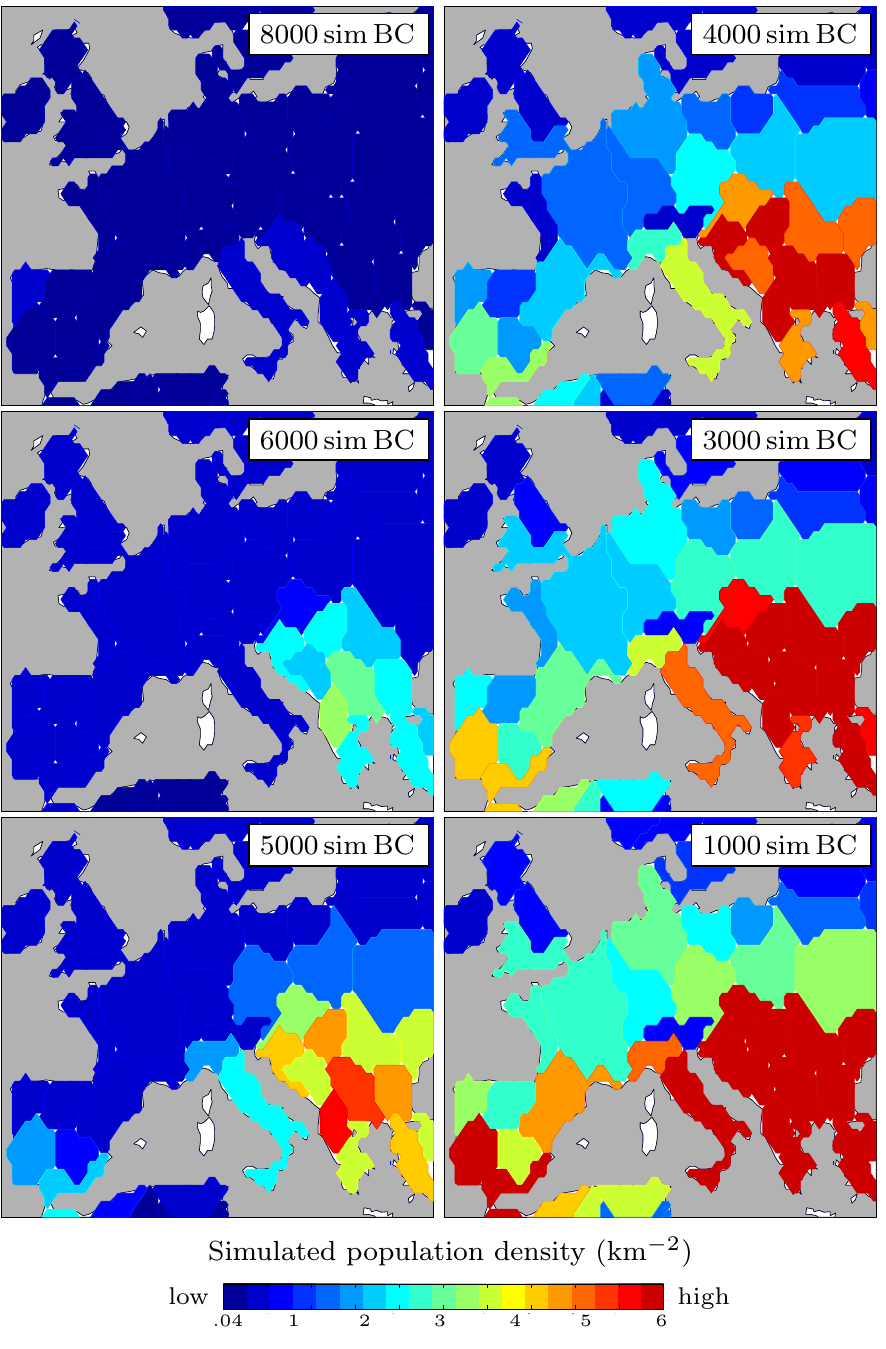}%
\ifreview\else}\hfil\parbox[c]{0.36\hsize}{\fi%
\caption{Population density (km$^{-2}$) in Europe between 8000\,{sim\,BC} and 1000\,{sim\,BC}, simulated with GLUES. }
\label{fig:map_europe_density}
\ifreview\else}\fi
\end{figure}

In Europe, the earliest simulated elevated population levels appear in Greece, Italy, Anatolia, and  the northwestern corner of the Iberian peninsula around 8000\,{sim\,BC} (Figure \ref{fig:map_europe_density}, time from top left to bottom right).  By 6000\,{sim\,BC}, this elevated population density of approximately one person per square kilometer is visible throughout Europe.  A dense population center with up to three people per square kilometer is focussed on northern Greece and extends throughout the Balkan and Anatolia.  At 5000\,{sim\,BC}, population intensifies further in these regions up to five people per square kilometer.  In Italy and on the southern Iberian peninsula population intensifies at this time.

By 4000\,{sim\,BC}, most of continental Europe exhibits population densities of a little above two people per square kilometer, with a densely populated region of five to six people per square kilometer in Greece, the Balkans, Anatolia and surrounding areas.  By 3000\,{sim\,BC}, Italy and the southern Iberian peninsula have experienced population boosts up to five people per square kilometer, much of central Europe is populated at a little less than three people per square kilometer.  By 1000\,{sim\,BC}, all along the Mediterranean coast population densities reach five to six people per square kilometer, and in most of the European woodland areas a population density of almost three people per square kilometer is supported.

\subsection{North American population expansion}

\begin{figure}
\centering
\ifreview\else\parbox[c]{0.53\hsize}{\fi%
\includegraphics[width=\ifreview0.8\else1\fi\hsize]{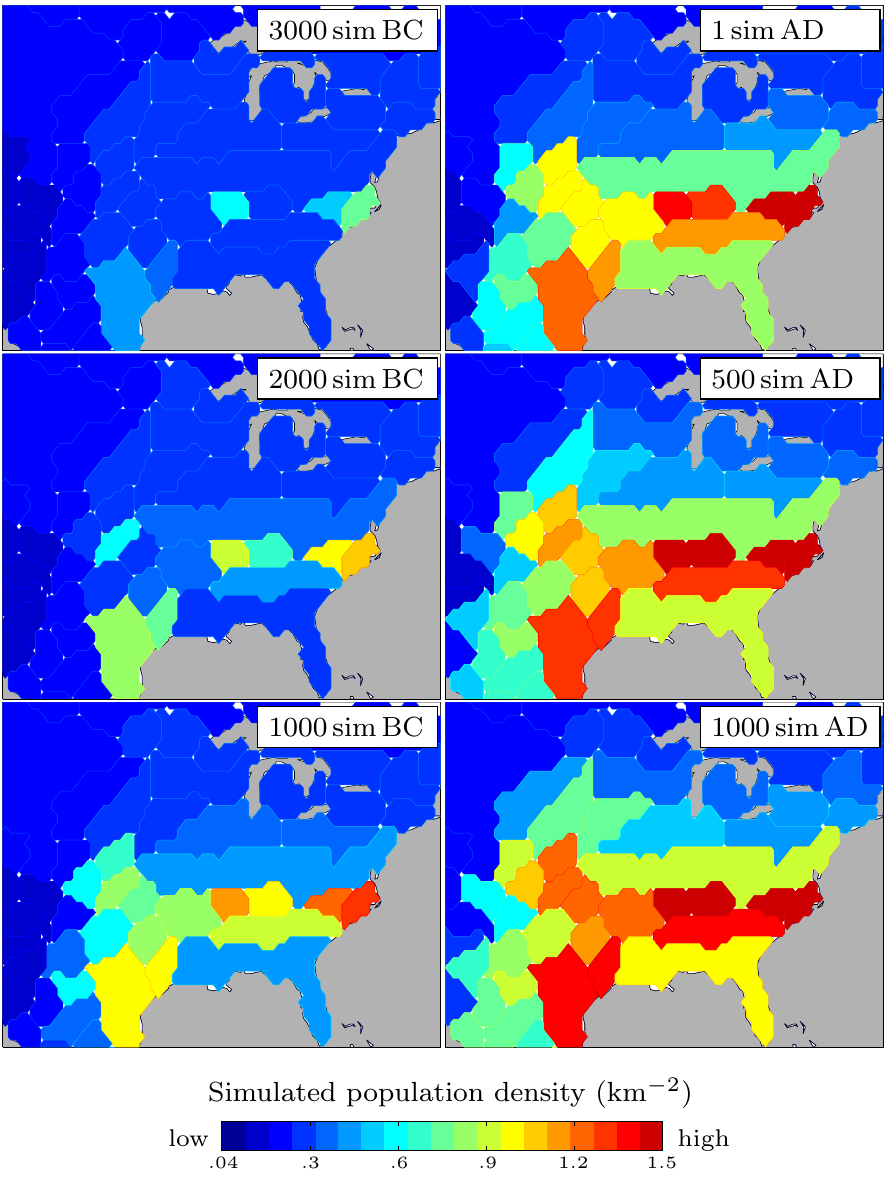}%
\ifreview\else}\hfil\parbox[c]{0.36\hsize}{\fi%
\caption{Population density (km$^{-2}$)  in North America between 3000\,{sim\,BC} (a) and 1000\,{sim\,AD} (f), simulated with GLUES.  Note the threefold scale difference to European density (Figure~\ref{fig:map_europe_density}, ranging up to 6\,\pskm\ in Europe and up to 2\,\pskm\ on the North American continent.}
\label{fig:map_american_density}
\ifreview\else}\fi
\end{figure}

In contrast to the development in Europe, the simulated emergence of high population density in North America, supported in part by agricultural or pastoral practice, occurs later in time and reaches only half of the maximum density attained in central Europe.  A first increase of population from the background level of 0.3~people per square kilometer population density is visible at 3000\,{sim\,BC} in the area of coastal Virginia and North Carolina (Figure \ref{fig:map_american_density}). The lowest population density is simulated along the Rocky Mountains with less then 0.1~people per square kilometer.   By 2000\,{sim\,BC}, the population density in these states reaches 0.8~people per square kilometer, population density is slightly elevated above the background level in parts of eastern Texas, Arkansas, Missouri, Tennessee, Mississippi, Alabama and Georgia.  By 1000\,sim\,BC, the population density in this southern Eastern North American region and in Texas reaches up to one person per square kilometer, with the highest density in a region on the Mississippi river between Arkansas, Missouri, Kentucky, and Tennessee.

From 1\,AD, the local population centers of the southern ENA join with eastern Texas to form  a large region of dense population extending from Mexico throughout the the southern ENA to the Atlantic coast.   Highest population density is simulated in the coastal states of North Carolina and Virginia with 1.1~people per square kilometer.  By 500\,sim\,AD population in this core area continues to increase, while in the coastal states to the south, the midwest states to the west and up the fourtieth parallel north of this core area population increases to 0.6~people per square kilometer.  By 1000\,sim\,{A.D} population increases in the entire ENA region south of the fourtieth parallel up to 1.5~people per square kilometer at the Atlantic coast and in the center, and up to 0.9 in the remaining part of ENA.  In the region north of the fourtieth parallel, low population densities of up to 0.4~people per square kilomter south of the Great Lakes (up to 0.35 in Canada) are hindcasted by the simulation.

\subsection{Trajectories to woodland farming}

\begin{figure}
\centering
\ifreview\else\parbox[c]{0.53\hsize}{\fi%
\includegraphics[width=\ifreview0.8\else1\fi\hsize]{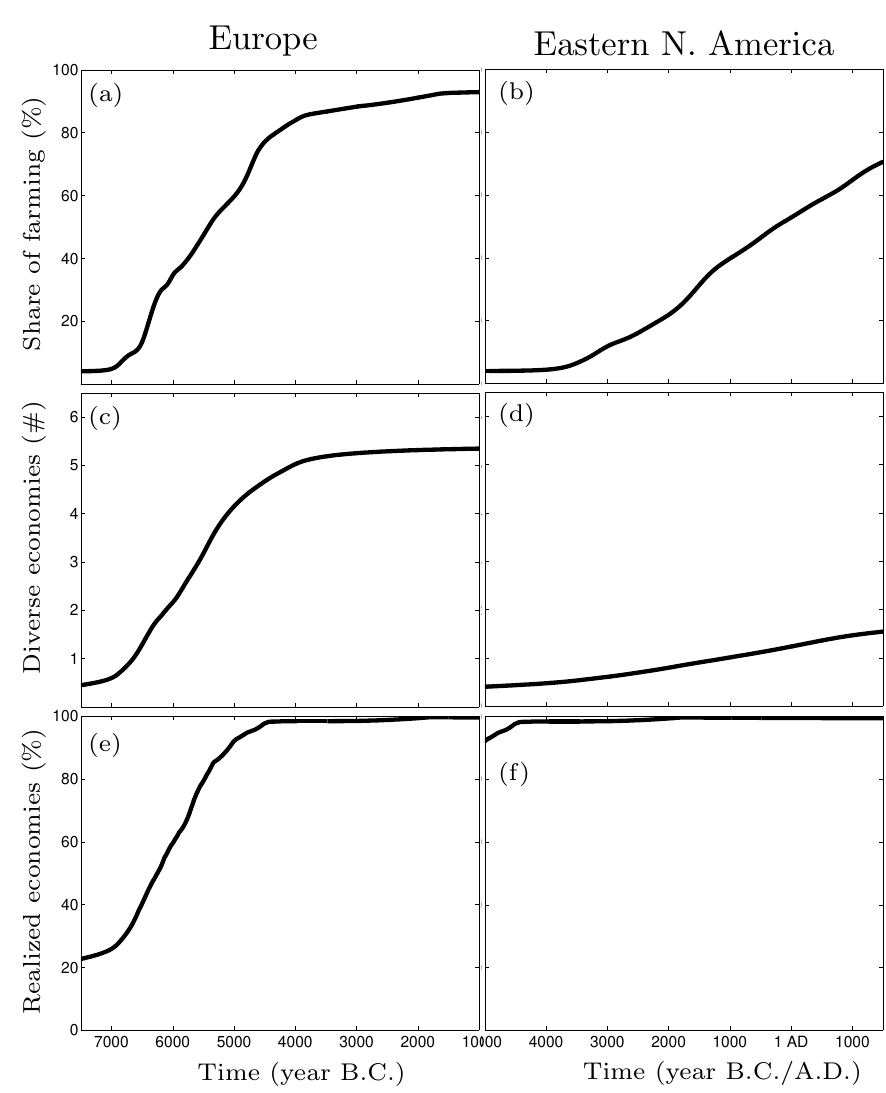}%
\ifreview\else}\hfil\parbox[c]{0.36\hsize}{\fi%
\caption{Trajectories of socio-technological development in European (left column, panel a,c,e) and Eastern North American (right column, panel b,d,f) woodlands simulated with GLUES and averaged over all regions in Europe and ENA, respectively.  From top to bottom: (a,b) fraction of agropastoral subsistence, (c,d) number of diverse economies, and (e,f) percentage of used economies relative to the natural potential.}
\label{fig:traj}
\ifreview\else}\fi
\end{figure}

Sociocultural trajectories to woodland farming are contrasted for Europe and the American East in Figure~\ref{fig:traj}.  This figure shows the development of the two socio-cultural variables fraction of agropastoralism (panels a, b) and economic diversity (panels c,d), and the realized fraction of the economic potential (panels e,f).  For both Europe and ENA, a 6500 year interval encompassing the transition to agriculture is displayed, 7500--1000\,sim\,BC for Europe, and 5000\,sim\,BC to 1500\,sim\,AD for ENA.  Most of the European transition takes place between 6500 and 4000\,sim\,BC; in this 2500 year interval 80\% of the transition occurs (panel a).  In contrast, the American transition starts after 4000\,sim\,BC and is not completed during pre-Columbian times, i.e., the duration of the transition is more than twice as long as for Europe (panel b).  Not only at the subcontinental level, but also at the individual simulation region level, the transition in the Americas takes more then twice as long (around 600 years) as the transition in most European regions (around 250 years).   

The middle panels in Figure \ref{fig:traj} show the trajectories of economic diversity---indicative of the number of domesticates in use.  There is a marked difference between Europe (panel c) and Eastern North America (panel d).  In Europe, the model simulates a gradual average increase of this quantity, resembling the increase of the importance of agriculture.  On average, five~diverse economies are in use by European farmers from 4000\,sim\,BC, however, this number stays approximately constant until the end of the simulation.  
From the same initial value of one half, the development in the Americas is much slower: from about 4000\,sim\,BC, the number of used diverse economies gradually increases but does not exceed 1.5 during the simulation.  The simulated large increase in agropastoral  activity in ENA relies on only a few different economies.

While it is clear from this and earlier studies that the different maximum value is a consequence of the different biogeographic setting, I find a marked difference between the two subcontinents  Europe and ENA in the relationship between actually used economies and the potentially available (from local or imported resources) economies; it is shown in panels e and f of Figure~\ref{fig:traj}, respectively.   The fraction of used economies increases in Europe from around 20\%--100\% before 4000\,sim\,BC.   The low initial fraction indicates that the abundant resources are hardly used and that the system is people-limited: there are not enough farmers to put into practice the economic potential during the Neolithic.  Only from the Chalcolithic period, the variety of resources is efficiently utilized in Europe.  
In North America, already at the beginning of the transition to agriculture around 4000\,sim\,BC the full spectrum of the economic potential of the landscape is utilized.  Here, the system is resource-limited, and the import of new economies drives the diversification of agricultural practice.  

\subsection{Woodland population development}

\begin{figure}
\centering
\ifreview\else\parbox[c]{0.53\hsize}{\fi%
\includegraphics[width=\ifreview0.8\else1\fi\hsize]{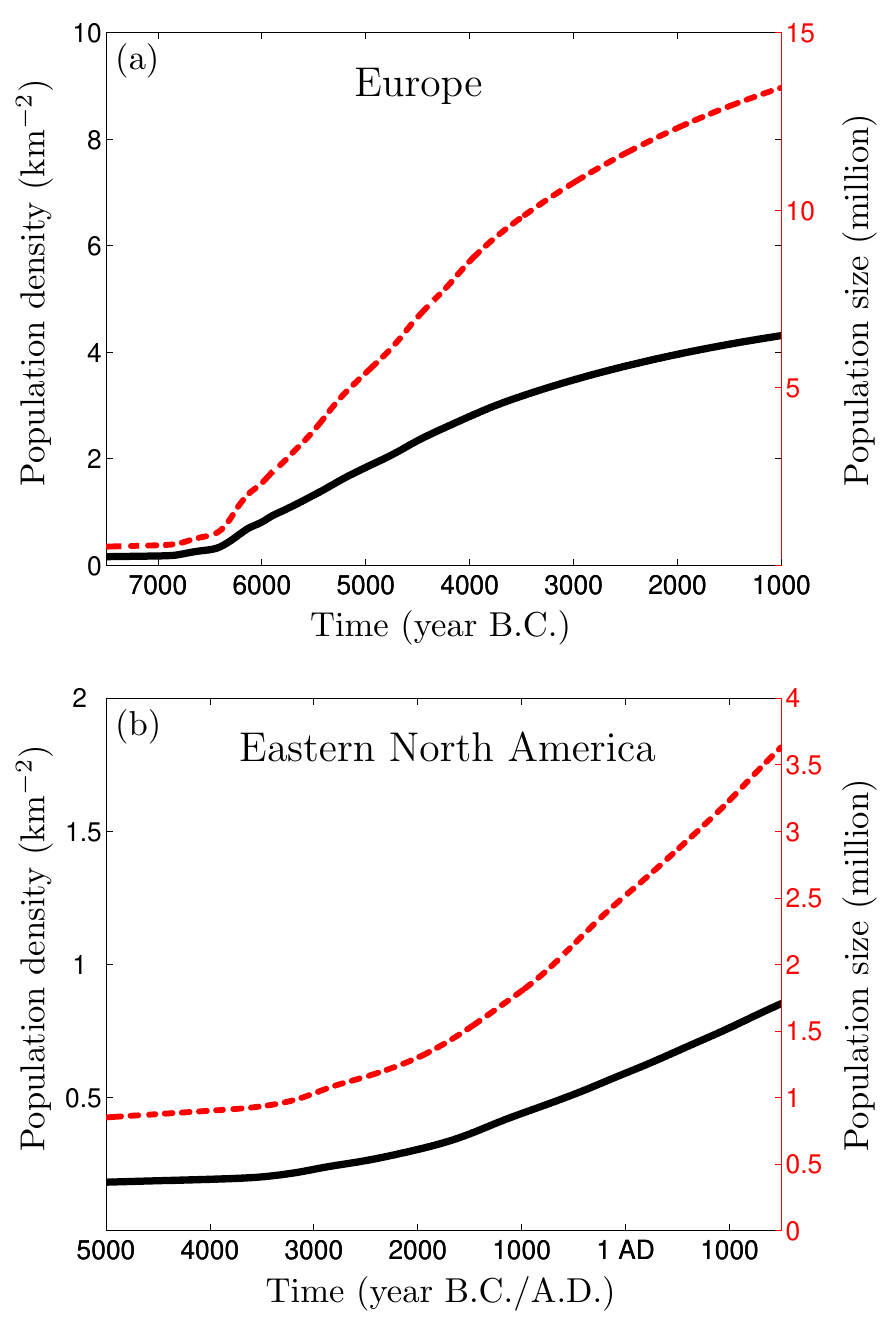}%
\ifreview\else}\hfil\parbox[c]{0.36\hsize}{\fi%
\caption{Trajectories of demographic development in European (top, panel a) and Eastern North American (bottom, panel b) woodlands simulated with GLUES and averaged over all regions in Europe and Eastern North America, respectively.  Both estimates of population density (solid line, left axis) and population size (broken line, right axis) are shown.  Note the time offset of 2500\,a between the two regions and the differing scales of the size estimates.}
\label{fig:trajpop}
\ifreview\else}\fi
\end{figure}

Average European population density (panel a) increased rapidly between 7000\,{sim\,BC} and 4000\,{sim\,BC} from 0.1~to 3~people per square kilometer.  Between  4000\,{sim\,BC} and 1000\,{sim\,BC}, the development is slower with another increase of 1.3~people per square kilometer.   On a total area of 3.1~million square kilometers, this population density translates to an effective population size of 13 million at 1000\,{sim\,BC}. Individual regions show evidence of diverging evolutions in different European regions, with population `explosions' in some areas of the southeast (e.g., Anatolia, Greece, Balkan), moderate population increases in the central areas of Europe and very low population density in the northern parts (cmp.~Figure~\ref{fig:map_europe_density}).

Pre-Columbian population in the American Woodland area is hindcasted by the model at around 3.5 million (Figure \ref{fig:trajpop}a).  Of these, one million existed already in the Archaic period; the increase during the Woodland and later period is 2.5~million.  The transition to agriculture and pastoralism is not completed at the end of the pre-Columbian period, when hunting and gathering were relevant on half the ENA Woodland area (Figure \ref{fig:trajpop}b).  The American trajectory to agriculture is characterized by an accelerating increase between 4000\,{sim\,BC} and 1\,{sim\,AD}, and a linear increase between 1\,{sim\,AD} and 1490\,{sim\,AD}.  Key features of the transition on both subcontinents are summarized in \reftab{tab:features}.

\section{Discussion}

\begin{table}
\begin{tabular}{|p{0.37\hsize}|p{0.3\hsize}|p{0.3\hsize}|} \hline
& Europe & Eastern North America \\ \hline
Transition start & 6500\,sim\,BC & 4000\,sim\,B.C \\
Transition duration & 2500 years & $>$5500 years \\
Local resource diversity potential  & 2.8 &   1.0 \\
Imported resource potential & 2.5 &  0.5 \\
Early limitation & people limited & resource limited \\
Peak population & 13 million @ 1000\,sim\,BC &  3.5 million @ 1490\,sim\,AD \\ \hline
\end{tabular}
\caption{Summary of key characteristics of the simulated transition to farming in Europe and Eastern North America.}
\label{tab:features}
\end{table}

\subsection{Neolithic package and mixed subsistence}

The instant availability of a whole array of diverse economies in Europe can be attributed to arrival of the large Neolithic package, which included, for example, emmer, einkorn, barley, goats, sheep, and pigs  \citep{Willcox2005}.  The rapid increase in people in the agropastoral sector is an indicator that not only the domesticates have been imported, but possibly also the people familiar with agropastoral life style.  This can be interpreted as an outplacement of the local hunting-gathering population and replacement by migrating farmers, consistent with the demic diffusion theories put forward by \citet{Ammerman1973}, \citet{Sokal1991}, or recently \citet{Haak2010}.  Alternatively, the fast conversion to agropastoralism can be viewed as fast adoption by resident foragers; this mechanism is supported by model simulations with the GLUES model by \citet{Lemmen2011}: the authors show that the transition to agriculture in Southeastern and Central Europe is consistent with both a demic or a cultural diffusion model, and that in either case local adoption played a major role.  

Much differently, domesticates in the Eastern North American Woodland were comprised of a a small but diverse array of local and imported crops.  Locally available were goosefoot, squash and sunflower---all part of the Eastern Agricultural Complex \citep{Delcourt1998}.  Gradually, local horticulture was supplemented with the foreign domesticates squash, maize, and beans.  These were often grown together in an assemblage termed the `Three Sisters', and originated from the Mexican founder centers.
The slow increase and the long transition duration in all socio-technological trajectories in the American Woodland point to cultural diffusion as the main mechanism of interregional exchange, and to slow adoption of new ways of subsistence.

\subsection{Global and Pre-Columbian Population}

The primary goal of this paper is not to discuss and provide a new estimate of pre-Columbian population in the Americas.  I find it instructive, however, to use estimates of population size published in the literature as cornerstones for evaluating the model performance with regards to simulating population.
The  global population estimates provided by GLUES are within the---admittedly large---range of  uncertainty of published studies on global prehistoric population \citep[e.g.][]{Coale1974,McEvedy1978,Biraben2003}.    At the subcontinental level, the GLUES estimate of ten million Europeans (Figure \ref{fig:traj}a) aligns perfectly with the regional data compiled by \citet[][]{McEvedy1978}  or with the regional compilation by \citet{Kaplan2010} for 1000\,BC.  Thereafter, with iron age technology, GLUES may underestimate the technological development and population density increases in Eurasia, because the efficiency gains by metal working and increased social stratification are not implemented in the cultural model yet.  For the Americas, Africa, and Australia, the simulations are valid until the beginning of colonialism, which is not implemented in the model yet.

For pre-Columbian America, the uncertainty within and between  population studies is very large. Published population estimates show more than an order of magnitude range between 10 and 150~million on the American continent; a value of around 60~million was assumed by \citet{Nevle2008}'s study on the impact of prehistoric population.    
Of these, probably only 10\% lived north of Mexico, while most of the population was concentrated in the Aztec and Incan empires, and in the Amazon basin \citep{Denevan1992}.  For the American Woodland area,  a rough estimate of four million individuals seems appropriate based on \citet{Nevle2008} and \citet{Denevan1992}.

Only recently, \citet{Peros2010} calculated postglacial and precontact population growth rates for North America based on the Canadian Archaeological Radiocarbon Database (CARD, \citealt{Morlan2005}).  Their calculation is a maximum pre-Columbian population consisting of 1.8--4.7 million individuals, with a best estimate of 2.5~million.   The GLUES model hindcasts 3.5~million individuals at 1490\,sim{\,AD} for Eastern North America (Figure \ref{fig:traj}d), for the whole of North America, the simulation yields 6.5~million people.  Given the uncertainties in obtaining the population estimates, GLUES provides a realistic population estimate for North America.

\subsection{Model critique} 
The good representation of pre-Columbian Eastern American population seems surprising, given the simplistic approach and parameter uncertainties associated with the socio-cultural model GLUES.  What the model does is to deterministically translate the geographical setting and vegetation resources into a potential for cultural evolution, with universal parameter settings for all world regions; then the cultural developments within each region and the interaction by trade and migration between regions produce estimates of demography, subsistence and technological advance. I interpret the success of this kind of simulation in the way that the large scale patterns of cultural evolution worldwide are based on very similar functional dependencies on biogeographic context and cultural interactions, despite the wide range of observed differences in cultural traits. 

The model could be improved even further if good chronologies were available for a multitude of archaeological sites and with concurrent clear interpretation.  For Europe, \citet{Lemmen2011} used the data compilation of 765~radiocarbon dates with Neolithic attribution from \citet{Pinhasi2005}; they found a model-data mismatch of $\pm$500 years on average.  For the American woodland, I here use a
 selection of  radiocarbon dates from CARD \citep{Morlan2005}:  3705 dates carry the designation `Woodland'; the central 95 percentile of these dates is confined to the period 1000\,BC to 1500\,AD, with maximum frequency in the 11th century AD.  When I apply this statistic to the GLUES model regions located within the area where radiocarbon dates with Woodland signature have been identified, the 23~model regions' timing of agropastoral onset is represented by  a broad distribution between 3500\,{sim\,BC} and 1490\,{sim\,AD}, with maximum occurrence between 2000\,{sim\,BC} and 1000\,{sim\,AD}.  For  ENA, there seems to be a model bias towards earlier dates.

Compilations of radiocarbon dates, however, are problematic on their own, due to the differences in the dated material, the differences in calibration procedure, different labs employing diverse quality checks and analysis methodology, and subjective choice of sites to include in the data base; not the least, the calibration itself introduces a dating uncertainty \citep[e.g.][]{Stuiver1998}.  The greater problem to overcome in a model-data comparison as done in this study is, however, the mismatch in the definition of the transition to agriculture.  In the simulation, a quantitative measure for the timing of the transition to agropastoralism is available, defined as the time (sim\,BC) when more than half of the food production is performed by agropastoralists.  This measure is not available in the archaeological record, where---amongst others---remains of crop or domestic animal taxa, agricultural tools, pictograms of agricultural technology give a hint of the existence, not the quantity, of agricultural activity.  When this is associated with a ceramic or stone tool characteristic, agricultural activity may be inferred for sites without direct evidence of agriculture.  There is no measure of farming activity from the archeological record which can be directly compared to the simulation results.

Using the regional patterns of the appearance of agropastoral evidence in the radiocarbon record and the model, and using indirect measures like population density or size, we can, however, make quantitative comparisons between data and models. In doing this, the most interesting finding should be the failure of a model to represent the observations: 
where the model hindcast deviates from the historic pathway, that is an indication of the need for explanation in the cultural realm. 
One such example could be the unusually (with respect to the natural capacity of the environment) low population density estimated for hunting-gathering societies in North America when compared to Europe.   This may have led the model--which includes a higher population according to the natural capacity---to simulate the introduction of agriculture in the ENA Woodland economy much earlier than has been observed (\reffig{fig:map_american_density}). Another example of a model shortcoming is the incomplete transition to agriculture in North England, Ireland or Scandinavia, which is observed in the radiocarbon record \citep[e.g.,][]{Price2003}, but not simulated by GLUES (\reffig{fig:map_europe_density}).  

The model presented here operates at the large level of a society.  Necessarily, it does not consider nor address individual  cultural choice and agency, i.e.\ of how people actively create and change the society they live in \citep{Dornan2002}.   Should the model be required to include agency at the collective level of a society, as suggested by \citet{Shanks1987}?  I believe not, because by explicitly excluding agency and simulating only the non-agency part of socio-technological evolution, the comparison of the model prediction and the historic record reveals where agency was important.  In other words, the model predictions 
should be interpreted as providing  `a historical null hypothesis. Its predictions can be taken as requiring no special explanation, and its failures can be taken as evidence of rare events that had significant and long-lived consequences' \citep{Ackland2007}.

The current model version lacks transport via sea routes and it has no representation of rivers.  The planned future implementation of sea connections would increase the realism of simulations on the Mediterranean islands, in Great Britain (which is connected to France by an artificial land bridge in the model) and Ireland.  While we could successfully 
simulate the Indus Valley transition to agriculture \citep{Lemmen2011harappa}, the implementation of rivers would make the model applicable to a river-based society like Egypt, where the moisture source is far from the agricultural area.

\section{Conclusion} 
I presented a global numerical model of regional transitions to agriculture.  This model is capable to realistically simulate a potential history which can then be compared to archaeological evidence.  The validity of the simulation is exemplified by the model prediction of realistic values for population density: ten million Europeans by 1000\,{sim\,BC}, and 6.5~million people on North America (of these 3.5~million in the Woodland area) by 1490\,{sim\,A.D}.

Despite the similar environmental setting of the American and European woodlands, and despite the common allochtonous origin of the major staple crops, the transition to agriculture and pastoralism proceeded very differently in these two regions (\reftab{tab:features}).  In Europe, an early and rapid adoption of the new lifestyle is simulated with the socio-technological simulator GLUES, consistent with the archaeological record.  A radial expanse from the Eastern Mediterranean with a speed comparable to the wave-of-advance model is realistically simulated, almost all of Europe has adopted agropastoralism by 4000\,{sim\,BC}. The time span of the transition is short and rather people-limited than resource limited.

If the same global parameter set is used to simulate the onset of agriculture in the American woodland, the model predicts a qualitatively different transition.  It is based on long-term coexistence of both hunting-gathering and agropastoral subsistence, on a mixture of local and foreign domesticates, transmitted by cultural diffusion rather than demic diffusion.  The transition can be characterized as resource-limited; it occurs late in time---starting around 2000\,{sim\,BC}---and has not completed by 1490.  

\paragraph{Acknowledgments.}
This study was partly funded by the German National Science Foundation (DFG priority project 1266 Interdynamik) and the PACES programme of the Helmholtz Association of German Research Centers.  Neither of the sponsors had any influence on the content or the submission decision of this manuscript. I thank K.\,W.~Wirtz for his contributions to and innovations in the GLUES model and fruitful comments on prior versions this manuscript.  I am grateful to D.\,G.~Anderson for his invitation to present a prior version of this paper at the 75$^\mathrm{th}$ annual meeting of the Society for American Archaeology (St.\ Louis, April 14--18, 2010) in the session ``Trajectories to Complexity in Woodland Environments: Eastern North America and Temperate Europe compared''.

\bibliographystyle{aeae}
\bibliography{journal-macros,glues}
\end{document}